\documentclass[12pt]{amsart}

\usepackage{xcolor}

\usepackage{amsmath,amsfonts,amsthm,amssymb,amsxtra}
\usepackage{tikz}
\usepackage{tikz-3dplot}

\usepackage{euscript}
\usepackage{multicol}
\usepackage{amsmath}
\usepackage{times}

\newtheorem{theorem}{Theorem}
\newtheorem{proposition}{Proposition}[section]
\newtheorem{corollary}{Corollary}[section]
\newtheorem{lemma}{Lemma}[section]

\newcommand{\beq}{\begin{equation}}
\newcommand{\eeq}{\end{equation}}

\def\ve{\varepsilon}

\def\l{\lambda}

\title[Absolutely continuous spectrum]{The rate of  accumulation of negative eigenvalues to zero and  the absolutely continuous spectrum}
\author{Oleg Safronov}

\address{Department of Mathematics and Statistics, University of North Carlolina at Charlotte,  9201 University City Blvd, Charlotte, NC 28223}

\begin{document}

\maketitle

\begin{abstract}
For a  bounded real-valued  function $V$ on ${\Bbb R}^d$, we consider two Schr\"odinger operators $H_+=-\Delta+V$  and  $H_-=-\Delta-V$.
We prove that if  the negative spectra $H_+$  and  $H_-$  are discrete and  the negative eigenvalues of $H_+$  and  $H_-$ tend  to zero
sufficiently fast, then
 the absolutely continuous spectra  cover the positive half-line  $[0,\infty)$.
\end{abstract}

\section{Main results}

Being discrete and being  continuous are two opposite properties  of a set in the  plane.
However, there are situations  in which  the fact  that one part of the set  is discrete implies that the other part is continuous.
The set    below  
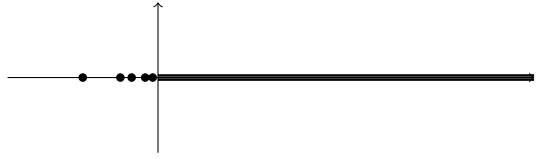
\begin{center}

\begin{tikzpicture}

\draw[thin,->] (5,-1) -- (5,1);

\draw[thick,-] (5,.03) -- (10,.03);

\draw[thick,-] (5,-.03) -- (10,-.03); \draw[thick,-] (5,0) -- (10,0);

\draw[thin,->] (3,0) -- (10,0);


\tikzstyle{every node}=[draw,shape=circle]

\draw[xshift=-1cm] (5.5,0) node[circle,fill,inner sep=1pt](a){};

\draw[xshift=-1cm] (5.65,0) node[circle,fill,inner sep=1pt](a){};

\draw[xshift=-1cm] (5.83,0) node[circle,fill,inner sep=1pt](a){};

\draw[xshift=-1cm] (5.93,0) node[circle,fill,inner sep=1pt](a){};

\draw[xshift=-1cm] (5,0) node[circle,fill,inner sep=1pt](a){};

\end{tikzpicture}

\end{center} 
has two parts: 
 the discrete part  (to  the left  of the vertical arrow),  and  
  the continuous  one (to the right of the arrow).
In general,  one  part  is not related  to   the other.   That is no  longer  true    if  
this picture   represents  the spectrum  of a Schr\"odinger operator!

 There is a relation  between the two parts of the spectrum. It  is particularly simple  if the potential $V(x)$ in  the Schr\"odinger equation is  bounded and negative.
In this case, if    the left part  of the spectrum is  discrete, then    the right  part  is continuous.  Moreover,  the continuous part coincides with the half-line $[0,\infty)$.
In   the general case, 
one has to consider  two  Schr\"odinger operators, one of which  is  obtained  from the other  by  flipping   the  sign of  the electric potential $V(x)$ at every point $x$.

\smallskip

\begin{theorem}\label{notmain} Let $V$  be a  real-valued  bounded measurable  function on ${\Bbb R}^d$.
If the negative spectra of the  two    Schr\"odinger operators  $H_+=-\Delta+V$ and $H_-=-\Delta-V$ are discrete,  then    both  spectra  contain every  point of the interval $[0,\infty)$.
\end{theorem}

\smallskip

 This theorem admits   mathematical assumptions   of the  form $V\in L^p_{loc}({\Bbb R}^d)$ that allow  usual  singularities  of $V$   appearing in physics (see \cite{KMS}).

\smallskip

 The rate of accumulation of eigenvalues  to zero   determines   certain
     properties   of  the positive  spectrum.
If the negative  eigenvalues tend to zero  sufficiently  fast,
we can talk about    absolute  continuity of the positive part.   Absolute continuity is    a mathematical notion  that is not easy to describe.
An absolutely continuous spectrum can  be seen in  a rainbow  in which  one color  is  consecutively  followed by another.
The colors change  from red to violet   so gradually and smoothly, that one  gets  an impression  that  this passage is   "absolutely continuous".

\smallskip

\begin{theorem}\label{main} Let $V$  be a  real-valued  bounded measurable  function on ${\Bbb R}^d$.
 Assume that the negative  spectra of both  operators  $H_+=-\Delta + V$ and $H_-=-\Delta-V$  
consist of isolated  eigenvalues $\{\l_j^+\}_{j=1}^\infty$  and $\{\l_j^-\}_{j=1}^\infty$ satisfying the condition
$$
\sum_j  |\l_j^+|^{1/2}+\sum_j  |\l_j^-|^{1/2}<\infty.
$$ 
Then the absolutely continuous spectrum of each of the two operators is essentially  supported on  the positive  half-line $[0,\infty)$.
\end{theorem}

\bigskip

The   last line  of the theorem should be  understood in the sense that the  density   of the spectrum   is positive almost everywhere on $[0,\infty)$. Namelly,  for  each $f\in L^2({\Bbb R}^d)$,  there is a unique non-negative measure $\mu_\pm$ on ${\Bbb R}$  having the   property
$$
\bigl((H_\pm-z)^{-1}f,f\bigr)=\int_{{\Bbb R}}\frac{d\mu_\pm(t)}{(t-z)},\qquad \forall z\in {\Bbb C}\setminus {\Bbb R}.
$$
The measure  $\mu_\pm$ is  said  to be  of maximal spectral type for $H_\pm$  provided   that any   condition of the form $\mu_\pm (\delta)=0$ 
 implies that the spectral projection    $E_{H_\pm}(\delta)$ of $H_\pm$  corresponding to the same Borel set $\delta\subset {\Bbb R}$  is    zero.
 By a  density of the spectrum we  mean the derivative of a spectral measure $\mu_\pm$  of the maximal spectral type. 
The  theorem  says that
$$
\mu'_\pm>0\qquad \text{almost everywhere on}\qquad [0,\infty).
$$

\bigskip

A  complete proof of   Theorem~\ref{notmain} can be  found in our joint paper \cite{KMS} with R. Killip and S. Molchanov.
The case $d=1$ of Theorem~\ref{main}  was  studied by D. Damanik and Ch. Remling. The corresponding proof for $d=1$  can  be  found  in \cite{DR}.

The main  goal of our  paper is to  present  a better  proof of Theorem~\ref{main} than  the unsatisfactory sketch given  in \cite{JFA}.
This proof is  different  from the one written  for $d=3$  in \cite{AHP} ,  because  it covers  all dimensions.

\section{Estimates of the potential}
The following theorem tells  us  that the rate of  accumilation of negative  eigenvalues to zero might determine  some  properties  of the potential.
\begin{theorem}\label{3.1} 
Let $W\geq 0$ be a  bounded  function on ${\Bbb R}^d$  having the property
$$
\int_{{\Bbb R}^d}\frac{W(x)}{|x|^{d-1}}dx<\infty.
$$
Let $V$  be a  real-valued   bounded  function on ${\Bbb R}^d$ and let $\lambda_j^\pm$ be   the negative  eigenvalues of  the Schr\"odinger operator $H_\pm=-\Delta+W\pm V$.
Suppose that
$$
\sum_j\Bigl(\sqrt{|\lambda_j^+|}+\sqrt{|\lambda_j^-|}\Bigr)<\infty.
$$
Then $V$ is representable in   the form
$$
V(x)=\tilde W(x)+{\rm div}\, A(x)+ |A(x)|^2,
$$
where  the vector potential $A:{\Bbb R}^d\rightarrow{\Bbb R}^d$  and   the function $\tilde W:{\Bbb R}^d\rightarrow{\Bbb R}$  satisfy the conditions
\begin{equation}\notag
\begin{split}
A\in L^\infty_{\rm loc}({\Bbb R}^d,{\Bbb R}^d)\cap  {\mathcal H}^1_{\rm loc}({\Bbb R}^d,{\Bbb R}^d),\quad \tilde W\in L^\infty_{\rm loc}({\Bbb R}^d),\\
\int_{{\Bbb R}^d}\frac{ (|\tilde W(x)|+|A(x)|^2)}{|x|^{d-1}}dx<\infty.\,\,\,\,\,\,\,\,\,\,\,\,\,\,\,\,\,\,\,\,
\end{split}
\end{equation}
\end{theorem}

\noindent {\bf Remark.}  The theorem does not  say  that  the functions $\tilde W$ and $A$   have to be  bounded or have to decay at infinity.

\bigskip

The next statement  can be  proved by integration  by parts.

\begin{lemma}\label{l3.1}
Let $\phi$ be a real-valued  bounded  function with  bounded derivatives of  first  order  defined  on a  domain $\Omega\subset{\Bbb R}^d$.  Suppose that   $\psi\in {\mathcal H}^2(\Omega)$  
is a real-valued  solution of
$$
-\Delta\psi+(W\pm V)\psi=\lambda \psi
$$
and  the product $\phi \psi$  vanishes  on the  boundary  of  the domain $\{a< |x|<b\}\subset \Omega$ with $a>0$. 
Then
$$
\int_{a<|x|<b}\Bigl(|\nabla (\phi\psi)|^2+(W\pm V)|\phi\psi|^2\Bigr)dx=\int_{a<|x|<b}\Bigl(|\nabla \phi|^2 \psi^2+\l |\phi\psi|^2\Bigr)dx.
$$
\end{lemma}

Before  stating a very important lemma, we introduce the notion   of the inner size (width)  $d(G)$ of a spherical layer $G=\{a\leq  |x|\leq b\}$ by setting it equal to $d(G)=b-a$.
For two spherical layers $\tilde G=\{\tilde a\leq |x|\leq  \tilde  b\}$  and $G=\{a\leq |x|\leq b\}$, we say that $G$ encloses $\tilde G$, if $\tilde b\leq a$.

By the   Schr\"odinger operator $-\Delta+W\pm V$ on a  domain $\Omega\subset {\Bbb R}^d$  we  always mean an operator with the Dirichlet  boundary conditions. We  will  sometimes  denote  these  operators by 
$H_+\Bigl|_\Omega$  and $H_-\Bigl|_\Omega$.  More often   we will  denote them by  $H_+$  and $H_-$, 
but in this case, we  will  provide a  verbal description  mentioning the domain $\Omega$.

\bigskip

\begin{lemma}\label{3.2}
Assume that  the  lowest  eigenvalue of $H_\pm$  on  the domain $\{a<|x|<b\}$ is  the number $-\gamma^2$  where $\gamma>0$. Suppose that $b-a\geq 6\gamma^{-1}$.
Then there is a  spherical layer $\Omega\subset \{a<|x|<b\}$  with $d(\Omega)=6\gamma^{-1}$ such that the lowest eigenvalue of $H_\pm$ on $\Omega$ is not  higher than $-\gamma^2/2$.
\end{lemma}

{\it Proof.}  Let $\psi$  be the real  eigenfunction  corresponding to the eigenvalue $-\gamma^2$
for the problem on the domain  $\{a<|x|<b\}$ with the Dirichlet   boundary conditions. Put $L=\gamma^{-1}$
and pick a  number $c>0$  giving  the maximum  to the function $f(c)=\int_{||x|-c|<L}|\psi|^2dx$  on the interval $[a,b]$.
Define $\phi$  as  by
$$
\phi(x)=\begin{cases}
1 \qquad {\rm if}\quad ||x|-c|<L,\\
0\qquad {\rm if}\quad ||x|-c|\geq 3L,\\
3/2-||x|-c|/(2L),\qquad {\rm otherwise}.
\end{cases}
$$
By the   choice of  the number $c$,
\begin{equation}\label{RD}
\int_{a<|x|<b} |\nabla \phi|^2\psi^2dx\leq \frac{\gamma^2}2\int_{a<|x|<b} | \phi \psi|^2dx.
\end{equation}
Indeed, $|\nabla \phi|$ vanishes  everywhere  except  for  the two sperical layers of width $2L$, where  it equals $\gamma/2$.
Consequently,
$$
\int_{a<|x|<b} |\nabla \phi|^2\psi^2dx\leq \frac{\gamma^2}2\int_{||x|-c|<L} | \psi|^2dx=\frac{\gamma^2}2\int_{||x|-c|<L}  | \phi \psi|^2dx.
$$
Therefore,  by Lemma~\ref{l3.1} and  the inequality  \eqref{RD},
$$
\int_{a<|x|<b}\Bigl(|\nabla (\phi\psi)|^2+(W\pm V)|\phi\psi|^2\Bigr)dx \leq - \frac{\gamma^2}2\int_{a<|x|<b} | \phi \psi|^2dx.
$$
That proved  the result with $\Omega$ defined as  the intersection of the support of $\phi$  with the layer $\{a<|x|<b\}$. If $d(\Omega)<6\gamma^{-1}$,
then we  enlarge $\Omega$  until its width  becomes equal to $6\gamma^{-1}$. The bottom of the spectrum  of the corresponding  operator will not  move up  in this process.
$\,\,\,\,\,\,\,\,\,\Box$
\bigskip

\begin{lemma}\label{3.4}
Let $V$ and $W\geq 0$ be two  real valued   bounded potentials on ${\Bbb R}^d$.
Let $H_\pm=-\Delta\pm V+W$  be two Schr\"odinger operators acting on $L^2({\Bbb R}^d)$. Suppose  that the negative spectra of   the operators $H_\pm$
are discrete and  consist of eigenvalues $\{ \lambda_j^\pm\}$  satisfying
$$
\sum_j\Bigl(\sqrt{|\lambda_j^+|}+\sqrt{|\lambda_j^-|}\Bigr)<\infty.
$$
Then there is a sequence  of spherical layers $\Omega_n=\{x\in{\Bbb R}^d:\,\,\,a_n\leq |x|\leq b_n\}$  and a monotone sequence of numbers $\epsilon_n>0$
having the properties:

(1) $\sum_n \epsilon_n^{1/2}<\infty$ and the  widths $d(\Omega_n)$ of  $\Omega_n$  are  estimated   by 
\begin{equation}
\label{42}
d(\Omega_n)\leq 42 \epsilon_n^{-1/2},\qquad \forall n>1.
\end{equation}

(2) $H_\pm\geq 0$  on the set ${\Bbb R}^d\setminus \cap_n \Omega_n$. Moreover,
\begin{equation}
\label{42b}
H_\pm\geq -\epsilon_n\quad \text{on}\quad \Omega_n\cup \bigl({\Bbb R}^d\setminus \cup_{j<n}\Omega_j\bigr),\qquad \forall n.
\end{equation}

(3) If $\Omega_j\cap \Omega_n\neq \emptyset$, then  the width of the intersection $\Omega_j\cap \Omega_n$
is bounded  below  by $6\epsilon_k^{-1/2}$
$$
d\Bigl(\Omega_j\cap \Omega_n\Bigr)\geq 6\epsilon_k^{-1/2},
$$
where $k=\min\{j,n\}$.

(4)  For each index $n$, there  are at most two sets  $\Omega_j$ intersecting $\Omega_n$ and 
$$
{\rm dist}\,\Bigl(\Omega_n,\cup_{m<j(n)}\Omega_m\Bigr)\geq 6 \epsilon_{j(n)}^{-1/2},
$$
where $j(n)$ is the smallest index $j<n$   for which  the intersection $\Omega_j\cap \Omega_n$ is not empty.

(5)  Any ball $B_r=\{x\in {\Bbb R}^d:\,\,|x|\leq r\}$ of a  finite radius $r>0$  intersects  only a  finite number of  sets $\Omega_j$.

\end{lemma}

{\it Proof.}  
We will construct  the sets $$\Omega_n=\{x\in {\Bbb R}^d:\,\,a_n\leq |x|\leq b_n\}$$ inductively.  We  will also   construct 
auxiliary  sets $\omega_n=\{x\in {\Bbb R}^d:\,\,\alpha_n\leq |x|\leq \beta_n\}\subset \Omega_n$  whose  description  will take a lot of space in this  proof.
First, set 
$$
\omega_0=\{x\in {\Bbb R}^d:\,\,|x|\leq 6\epsilon_0^{-1/2}\}\qquad \text{and}\quad \Omega_0=\{x\in {\Bbb R}^d:\,\,|x|\leq 12\epsilon_0^{-1/2}\}
$$ where $-\epsilon_0$ is    the the lowest of the  eigenvalues $\{\lambda_j^\pm\}$.

\bigskip

Suppose the sets $\omega_n\subset \Omega_n$  and the numbers $\epsilon_n$  are already constructed  for   $n<N$. Consider the set
$$
S={\Bbb R}^d\setminus \cup_{n<N}\Omega_n
$$
and define $-\epsilon_N$  as the lowest of the eigenvalues of $H_+$ and $H_-$ on $S$.  By construction,
$$
\epsilon_j\geq \epsilon_{j+1}.
$$
Define $\omega\subset S$ to be the spherical layer on which  one of the operators $H_\pm$  has  spectrum below $-\epsilon_n/2$, i.e.
$$
{\rm inf}\,\sigma\Bigl(H_\pm\Bigl|_\omega\Bigr)\leq -\epsilon_N/2\quad \text{either for + or -},
$$
while the width of $\omega$  is not  larger than $L=6\epsilon_N^{-1/2}$.  We assume that  one can not enlarge $\omega$  preserving the properties  described  above.
The existence of this set is  proved in Lemma~\ref{3.2}.

\bigskip

Let    $\alpha$  and $\beta$ be the  non-negative numbers  defined by
$$\omega=\{x\in {\Bbb R}^d:\,\,\alpha\leq |x|\leq \beta\}.$$
Choose the index $l$  so that   $a_l$  is the smallest  of the numbers $\{a_n\}_{n<N}$  having the property
$$
\beta \leq a_n.
$$
After that, choose the index $k$ so that $b_k$ is  the largest of the numbers  $\{b_n\}_{n<N}$ having the property $$b_n\leq \alpha.$$
Note that the number $l$  might not exist. However, the case  where $l$  does not exist can be dealt with   as if $a_l$  was infinite.

{\bf Case 1}.   If $a_l-b_k<2 \max\{L_-,L_+\}$  where $L_-=6\epsilon_k^{-1/2}$  and $L_+=6\epsilon_l^{-1/2}$, then we  replace $\Omega_k$  and $\Omega_l$  by  two larger sets
so that  the  width of the intersection  will be  equal to
$$
d(\Omega_k\cap\Omega_l)=\min\{L_-,L_+\}
$$
For instance, if $L_-\leq L_+$, then we replace $\Omega_k$  by $\{a_k\leq |x|\leq b_k+L_-\}$  and   replace $\Omega_l$  by $\{b_k\leq |x|\leq b_l\}$.
This operation   would not  change   the property
$$
H_\pm\Bigl|_{\Omega_n}\geq -\epsilon_n\quad \text{for}\quad n<N,
$$
because  of  the claim 2) of the lemma.

After we  redefine  the two sets  $\Omega_k$ and $\Omega_l$, we start the process over with a new   collection   of the sets $\{\Omega_n\}_{n<N}$.

\medskip

{\bf Case 2}.  If  both $a_l-\beta>L$  and $\alpha-b_k>L$, then  we  set
$$
\Omega_N=\{x\in {\Bbb R}^d:\,\, \alpha-L\leq |x|\leq \beta+L\}
$$
and $\omega_N=\omega$.

\medskip

{\bf Case 3}. If $a_l-b_k\geq 2 \max\{L_-,L_+\}$, but $\alpha-b_k\leq L$ and $a_l-\beta\leq L$, then we set $\omega_N=\omega$,
$$
\Omega_N=\{x\in {\Bbb R}^d:\,\,\,b_k\leq |x|\leq a_l\},
$$
and we replace $\Omega_k$  and $\Omega_l$  by  the sets $$\{x\in {\Bbb R}^d:\,\,\,a_k\leq |x|\leq b_k+L_-\}\quad \text{and }\quad \{x\in {\Bbb R}^d:\,\,\,a_l-L_+\leq |x|\leq b_l\}$$  correspondingly.

\medskip

{\bf Case 4}.  Finally, consider the case  where $a_l-b_k\geq 2 \max\{L_-,L_+\}$, but only one of the numbers $\alpha-b_k$ and $a_l-\beta$  is not larger than $L$.
Let us assume that $\alpha-b_k\leq L$ but $a_l-\beta>L$. In this case, we set $\omega_N=\omega$,
$$
\Omega_N=\{x\in {\Bbb R}^d:\,\,\,\,b_k\leq |x|\leq \beta+L\},
$$ 
and we replace $\Omega_k$  by  the set $\{x\in {\Bbb R}^d:\,\,\,\,\, a_k\leq |x|\leq b_k+L_-\}$. 

\medskip

We see that initially the  width of $\Omega_N$  does not exceed $3L$. However, we might change $\Omega_N$  by $2L$ at the next  step of the process. 
 Since the number  of the steps at  which  one set  $\Omega_n$  can be changed is at most two, the width  of  $\Omega_n$  does not exceed $42\epsilon_n^{-1/2}$. That proves \eqref{42}.

Since the set $\Omega_N\cup\bigl({\Bbb R}^d\setminus\cup_{j<N}\Omega_j\bigr)$  is contained in $S$, we  the relation \eqref{42b} holds   for $n=N$.  Therefore it  holds  for any $n$
after  the   construction of the sets $\Omega_n$  is completed.

Obviously, we  extended  the sets $\Omega_n$  so that  the claim 3) holds. Since $\omega_k$ and $\omega_l$  have not been changed, the are at least  distance $L_-$ and $L_+$ apart  from $\Omega_N$.
Since $L_\pm\geq 6 \epsilon_{j(N)}^{-1/2}$, we obtain  the clam 4)  is  true.

The  sets $\omega_n$  are disjoint and one of the operators $H_\pm$   on $\omega_n$  has an eigenvalue below $-\epsilon_n/2$  for $n\geq 1$. Consequently,
$$
\sum_{n=1}^\infty\epsilon_n^{1/2}\leq \sqrt2\sum_n\Bigl(\sqrt{|\lambda_n^+|}+\sqrt{|\lambda_n^-|}\Bigr).
$$

It is also clear that a  ball $B_r$  of  finite radius $r>0$ can intersect only a  finite  number of the disjoint sets $\omega_n$. Otherwise the spectrum of  one of the operators $H_\pm\Bigl|_{B_r}$
would accumulate  to zero, which  can  never  occur on a bounded domain  due to one of Sobolev's embedding theorems. This implies the fifth  claim  of the lemma.

The fact  that,  for each $N$,
$$
{\Bbb R}^d\setminus \cup_n \Omega_n  \subset {\Bbb R}^d\setminus \cup_{n<N} \Omega_n
$$
implies that $H_\pm\geq -\epsilon_N$ on  ${\Bbb R}^d\setminus \cup_n \Omega_n$.  Consequently,  $H_\pm\geq 0$ on  ${\Bbb R}^d\setminus \cup_n \Omega_n$. $\,\,\,\,\,\,\,\,\,\,\,\,\Box$

\bigskip

\bigskip

Lemma~\ref{3.4}  allows one to estimate  the potential $V$ on the union $\cup_n\Omega_n$. However, these sets   might not  cover  the whole space ${\Bbb R}^d$,
so we have to consider the case
$$
{\Bbb R}^d\setminus \cup_n \Omega_n\neq \emptyset.
$$

\begin{lemma}\label{3.5}  Enlarging  some of the sets $\Omega_n$  from Lemma~\ref{3.4}, one can  achieve  that
$$
{\Bbb R}^d=\Bigl(\cup_n\Omega_n\Bigr)\cup \Bigl(\cup_n\Lambda_n\Bigr)
$$
where $\Lambda_n=\{x\in {\Bbb R}^d:\,\,\,\, \alpha_n<|x|<\beta_n\}$ are  spherical layers with the properties:

(1)  both operators $H_+$ and $H_-$ are positive on $\Lambda_n$

(2) each bounded layer $\Lambda_m$ intersects  exactly two sets $\Omega_n$

(3) if $\Lambda_n$  intersects $\Omega_{n_1}$  and $\Omega_{n_2}$, then
$$
d(\Lambda_n)\geq 6\epsilon_{n_1}^{-1/2}+6\epsilon_{n_2}^{-1/2},
$$
and
$$
d(\Lambda_n\cap \Omega_{n_j})=6\epsilon_{n_1}^{-1/2},\quad j=1,2,
$$
where $d(G)$ denotes the  width of $G$.

(4)  all claims of Lemma~\ref{3.4} hold  for the sets $\Omega_n$  except  for inequality \eqref{42}  which should be   replaced  by
\begin{equation}\label{44}
d(\Omega_n)\leq 67 \epsilon^{-1/2}_n,\qquad \forall n>1.
\end{equation}

\end{lemma}

{\it Proof.} Let  the collection of sets  $\{\Omega_n\}$ be the same as in Lemma~\ref{3.4}. The set ${\Bbb R}^d\setminus \cup_n \Omega_n$  is a  disjoint  union of  spherical layers on which   both operators $H_\pm$ are positive. 
If a  spherical layer $\Lambda$  is a   connected  component of  ${\Bbb R}^d\setminus \cup_n \Omega_n$ then there  are  two sets $\Omega_{n_1}$ and  $\Omega_{n_2}$  whose  boundaries intersect the boundary of 
$\Lambda$. In this case,  the width of $\Lambda$ should be  compared with $6\epsilon_{n_1}^{-1/2}+6\epsilon_{n_2}^{-1/2}$. If  $d(\Lambda)$ is  smaller than this number, we
enlarge  $\Omega_{n_1}$ and  $\Omega_{n_2}$  so that the gap between them will disappear. 
For instance, if $d(\Lambda)<6\epsilon_{n_1}^{-1/2}+6\epsilon_{n_2}^{-1/2}$, and $\epsilon_{n_1}^{-1/2}\leq \epsilon_{n_2}^{-1/2}$, then we  replace $\Omega_{n_2}$ by the union 
$\Omega_{n_2}\cup\bar\Lambda$  and give the piece  of width $6\epsilon_{n_1}^{-1/2}$ to the set $\Omega_{n_1}$.
 Otherwise,  if  if $d(\Lambda)\geq 6\epsilon_{n_1}^{-1/2}+6\epsilon_{n_2}^{-1/2}$,  we keep $\Lambda$ as  a member of the collection $\{\Lambda_n\}$. In this  case, we  enlarge both sets
$\Omega_{n_1}$ and  $\Omega_{n_2}$   giving them the pieces of $\Lambda$ of the width $6\epsilon_{n_1}^{-1/2}$  and $6\epsilon_{n_2}^{-1/2}$, correspondingly.

Since the width of $\Omega_n$  in this process  might  change at  most by $24\epsilon_n^{-1/2}$, we obtain  the inequality \eqref{44}. $\,\,\,\,\,\,\,\,\,\,\,\Box$

\bigskip

In order to obtain   the required estimates  of the potential $V$ we  need the following elementary statement.

\begin{lemma}\label{3.3}
Let  both  $H_+\geq -\gamma^2$ and $H_+\geq -\gamma^2$ on a  bounded  spherical layer
$\Omega=\{a<|x|<b\}$, $a>0$. Then $W+V+\gamma^2={\rm div}\,A+|A|^2$ on $\Omega$, where
the  vector potential $A\in L^\infty_{\rm loc}(\Omega;{\Bbb R}^d)\cap {\mathcal H}^1_{\rm loc}(\Omega;{\Bbb R}^d)$  satisfies the estimate
\begin{equation}\label{41}\begin{split}
\frac12\int_{a<|x|<b}|\phi|^2|A(x)|^2dx\leq \\
\gamma^2\int_{a<|x|<b}|\phi|^2dx+\int_{a<|x|<b}W|\phi|^2dx+3\int_{a<|x|<b}|\nabla\phi|^2dx.
\end{split}
\end{equation}
for any function $\phi\in C^\infty_0(\Omega)$.
\end{lemma}

{\it Proof.}  Let $u$ be  a  positive  solution  of the equation $-\Delta+(W+V)u=-\gamma^2 u$. Then $A=u^{-1}\nabla u$ is a vector potential
obeying
$$
W+V=-\gamma^2+{\rm div}\, A+|A|^2 \quad \text{on}\quad \Omega.
$$
This  step is  justified in my   paper \cite{22}. Now, the condition  $H_+\geq -\gamma^2$  can be written in the form
$$
\int_{a<|x|<b}\Bigl(|\nabla \phi|^2+(W-V)|\phi|^2dx\Bigr)\geq -\gamma^2\int_{a<|x|<b}|\phi|^2dx.
$$
The latter leads to  the inequality \eqref{41}  due to the  estimate
$$
\int_{a<|x|<b}{\rm div}\, A |\phi|^2dx  \leq \frac12\int_{a<|x|<b}|A |^2 |\phi|^2dx+2\int_{a<|x|<b} |\nabla \phi|^2dx.
$$
The proof is completed. $\,\,\,\,\,\,\,\,\,\,\,\,\,\,\,\Box$

\bigskip

Since  the  functions $\phi$ in Lemma~\ref{3.3} must vanish at the boundary of $\Omega$, this lemma  allows  one to estimate
$A$  only  inside  the domain.

\begin{corollary}\label{3.6}
Let  both  $H_+\geq -\gamma^2$ and $H_+\geq -\gamma^2$ on a  bounded  spherical layer
$\Omega=\{a<|x|<b\}$, where  $a,\,\gamma>0$  and $b-a\leq 67/\gamma$. Let also
$$\tilde\Omega=\{\tilde a<|x|<\tilde b\},$$ where $a<\tilde a<\tilde b< b$.
Then \begin{equation}\label{45}W+V+\gamma^2={\rm div}\,A+|A|^2\end{equation}
 on $\Omega$, where
the  vector potential $A\in L^\infty_{\rm loc}(\Omega;{\Bbb R}^d)\cap {\mathcal H}^1_{\rm loc}(\Omega;{\Bbb R}^d)$  satisfies the estimate
\begin{equation}\label{41b}\begin{split}
\frac12\int_{\tilde \Omega}|A(x)|^2|x|^{1-d}dx\leq \\
67\gamma+\int_{\Omega}\bigl(W+6|x|^{-2}\bigr)|x|^{1-d}dx+6\Bigl((\tilde a-a)^{-1})+(b-\tilde b)^{-1}\Bigr).
\end{split}
\end{equation}
\end{corollary}
{\it Proof.} The  inequality \eqref{41b}  follows  from \eqref{41} in which one has to   set $\phi(x)=\theta(|x|) |x|^{(1-d)/2}$,
where $\theta$   is a continuous  function on ${\Bbb R}$ defined by
$$
\theta(t)=\begin{cases}
0,\qquad \text{if}\qquad t\notin[a,b];\\
1,\qquad \text{if}\qquad t\in[\tilde a,\tilde b];\\
\text{ is linear  on }  [a,\tilde a];\\
\text{ is linear  on }  [\tilde b, b].
\end{cases}
$$
The proof is completed. $\,\,\,\,\,\,\Box$

\bigskip

Obviously, Corollary~\ref{3.6} holds  for $\gamma=0$. 
\begin{corollary}\label{3.7}
Let  both  $H_+\geq 0$ and $H_+\geq 0$ on a  bounded  spherical layer
$\Lambda=\{\alpha<|x|<\beta\}$, where  $\alpha>0$. Let also
$$\tilde\Lambda=\{\tilde \alpha<|x|<\tilde \beta\},$$ where $\alpha<\tilde \alpha<\tilde \beta< \beta$.
Then \begin{equation}\label{45c}W+V={\rm div}\,A+|A|^2\end{equation}
 on $\Lambda$, where
the  vector potential $A\in L^\infty_{\rm loc}(\Lambda;{\Bbb R}^d)\cap {\mathcal H}^1_{\rm loc}(\Lambda;{\Bbb R}^d)$  satisfies the estimate
\begin{equation}\label{41c}\begin{split}
\frac12\int_{\tilde \Lambda}|A(x)|^2|x|^{1-d}dx\leq \\
\int_{\Lambda}\bigl(W+6|x|^{-2}\bigr)|x|^{1-d}dx+6\Bigl((\tilde \alpha-\alpha)^{-1})+(\beta-\tilde \beta)^{-1}\Bigr).
\end{split}
\end{equation}
\end{corollary}

We can now  use the  information obtained in the two preceding corollaries to prove the following  statement.

\begin{lemma}\label{3.8}
Let $V$ and $W\geq 0$ be two  real valued   bounded potentials on ${\Bbb R}^d$.  Assume that
$$
\int_{{\Bbb R}^d}\frac{W}{|x|^{d-1}}dx<\infty.
$$ Suppose  that the negative spectra of   the operators $H_\pm=-\Delta\pm V+W$ 
are discrete and  consist of eigenvalues $\{ \lambda_j^\pm\}$ satisfying
$$
\sum_j\Bigl(\sqrt{|\lambda_j^+|}+\sqrt{|\lambda_j^-|}\Bigr)<\infty.
$$
Let $\Omega_n$, $\Lambda_n$  and $\epsilon_n$  be the same as  in Lemma~\ref{3.5}. Assume that  $\Omega_n\subset\{|x|\geq 6\epsilon^{-1/2}_0\}$  for all $n\geq 1$.
Then  there is a sequence of ${\mathcal H}^1$-functions $\phi_n\geq 0$  supported inside $\Omega_n$ and a sequence of ${\mathcal H}^1$-functions $\psi_n\geq 0$  supported inside $\Lambda_n$
such that
\begin{equation}\label{460}
\sum_n\phi_n(x)+\sum_n\psi_n(x)=1,
\end{equation}
\begin{equation}\label{461}
\sum_n \int_{{\Bbb R}^d}\bigl(|\nabla \phi_n(x)|^2+|\nabla\psi_n(x)|^2\bigr)|x|^{1-d}dx\leq 72\sum_n \epsilon_n^{1/2}
\end{equation}
Moreover, one can find  vector potentials $A_n$  and $\tilde A_n$  such that
\begin{equation}\label{46}
V+W+\epsilon_n={\rm div}\, A_n+|A_n|^2\quad \text{on}\quad \Omega_n,\qquad V+W={\rm div}\,\tilde  A_n+|\tilde A_n|^2\quad \text{on}\quad \Lambda_n.
\end{equation}
and 
\begin{equation}
\begin{split}
\frac12\sum_{n= 1}^\infty \Bigl(    \int_{{\rm supp}\,\phi_n}|A_n|^2|x|^{1-d}dx  +\int_{{\rm supp}\,\psi_n}  |\tilde A_n|^2|x|^{1-d}dx   \Bigr)\leq \\
(|{\Bbb S}_d|+500)\sum_n\epsilon_n^{1/2}+\int_{{\Bbb R}^d}\frac{W}{|x|^{d-1}}dx,
\end{split}
\end{equation}
where $|{\Bbb S}_d|$ is the area of the unit  sphere in ${\Bbb R}^d$.
\end{lemma}

{\it Proof.}  According to Lemma~\ref{3.4},  the  width of a non-empty set of the form  $\Omega_j\cap \Omega_n\neq \emptyset$
is bounded  from below  by $6\epsilon_k^{-1/2}$, where $k=\min\{j,n\}$. Also, according  to Lemma~\ref{3.5},  if $\Lambda_j\cap \Omega_n\neq \emptyset$, then the  width of the intersection
$\Lambda_j\cap \Omega_n$  is  not less than $6\epsilon_n^{-1/2}$.  Let  $$\{ r_n<|x|<R_n\}$$ be the enumeration of  the interiors of all such intersections  that has  the property $R_n\leq r_{n+1}$ for all $n$. Define the  functions  $\theta_n$ 
so that  they are  continuous on ${\Bbb R}$ and are  linear on the middle thirds $$\Bigl[ r_n+\frac{(R_n-r_n)}3,R_n-\frac{(R_n-r_n)}3\Bigr]\quad \text{  and } \quad \Bigl[ r_{n+1}+\frac{(R_{n+1}-r_{n+1})}3,R_{n+1}-\frac{(R_{n+1}-r_{n+1})}3\Bigr]$$ of the intervals $$[ r_n,R_n]\quad \text{  and } \quad[ r_{n+1},R_{n+1}],$$ correspondingly.  We define  $\theta_n$ to be identically   zero outside of $$[ r_n+\frac{(R_n-r_n)}3, R_{n+1}-\frac{(R_{n+1}-r_{n+1})}3].$$ Finally, we  define $\theta_n$  to  be identically equal to one
on the interval $$[ R_n-\frac{(R_n-r_n)}3,   r_{n+1}+\frac{(R_{n+1}-r_{n+1})}3].$$

Now  for  each index $n$, we set $\phi_n(x)=\theta_j(|x|)$, where  $j$  is the index  for which  the support of the function  $\theta_j(|\cdot|)$ is contained in $\Omega_n$.
Also, for  each index $n$, we set $\psi_n(x)=\theta_l(|x|)$, where  $l$  is the index  for which  the support of the function  $\theta_l(|\cdot|)$ is contained in $\Lambda_n$.

Observe  that
$$
\int_{\Omega_j\cap \Omega_n} |\nabla \phi_n|^2 |x|^{1-d}dx\leq 18 \epsilon_k^{1/2},\qquad k=\min\{n,j\}.
$$
$$
\int_{\Lambda_j\cap \Omega_n} |\nabla \phi_n|^2 |x|^{1-d}dx\leq 18 \epsilon_n^{1/2}.
$$
These relations imply \eqref{461}.
Moreover, $\phi_n+\phi_j=1$  on the set $\Omega_j\cap \Omega_n$  and  $\phi_n+\psi_j=1$  on the set $\Lambda_j\cap \Omega_n$. The latter  properties imply \eqref{460}.

The representations \eqref{46} as well as   the integral estimates   for $A_n$ and $\tilde A_n$  follow  from Corollaries~
\ref{3.6} and ~\ref{3.7}, because $H_\pm\geq -\epsilon_n$  on $\Omega_n$, and    both operators $H_\pm$ are positive on $\Lambda_n$. 
We  also use the  fact that 
$$
6\int_{|x|>6\epsilon_0^{-1/2}} |x|^{-2}|x|^{1-d}dx=|{\Bbb S}_d|\epsilon_0^{1/2}.
$$
$\,\,\,\,\,\,\,\,\,\Box$

\bigskip

{\it The end of the proof of Theorem~\ref{3.1}}.   Let us  
define
$$
A=\sum_{n=1}^\infty\bigl(\phi_n A_n+\psi_n \tilde A_n\bigr),\quad p(x)=-\sum_{n=1}^\infty\epsilon_n\phi_n(x),\quad V_1=p+{\rm div}\, A +|A|^2.
$$
Note that 
\begin{equation}\label{47}\begin{split}
\int_{{\Bbb R}^d}|p(x)|\,|x|^{1-d}dx\leq 42\sum_n \epsilon_n^{1/2}<\infty, \quad
\int_{{\Bbb R}^d}|A(x)|^2\,|x|^{1-d}dx\leq \\ 2\sum_{n=1}^\infty  \Bigl(\int_{{\rm supp }\, \phi_n}|A_n(x)|^2\,|x|^{1-d}dx+ \int_{{\rm supp }\, \psi_n}|\tilde A_n(x)|^2\,|x|^{1-d}dx\Bigr)<\infty.
\end{split}
\end{equation}
The relations  \eqref{46}  imply 
$$
\phi_n(V+W+\epsilon_n)=\phi_n ({\rm div}\, A_n+|A_n|^2),\qquad \psi_n(V+W)=\psi_n ({\rm div}\, \tilde A_n+|\tilde A_n|^2),
$$
Taking the sum over all $n$  and  using the property that $\{\phi_n\}$ and $\{\psi_n\}$  is  a partition of the unity, we obtain  the relation
$$
V+W-p=\sum_{n=0}^\infty \phi_n ({\rm div}\, A_n+|A_n|^2)+\sum_{n=0}^\infty\psi_n ({\rm div}\, \tilde A_n+|\tilde A_n|^2).
$$
Consequently,
$$
V+W=V_1 - \sum_{n=0}^\infty ( A_n \nabla \phi_n + \tilde A_n\nabla \psi_n )-|A|^2+\sum_{n=0}^\infty  ( \phi_n |A_n|^2+\psi_n|\tilde A_n|^2).
$$
This representation   implies that
\begin{equation}\label{48}
\int_{{\Bbb R}^d}|V+W-V_1|\,|x|^{1-d}dx<\infty,
\end{equation}
because  the  gradients of $\phi_n$  and $\psi_n$  obey  the condition \eqref{461}.

It  remains to  set $\tilde W=V-V_1+p$. Then
$V=\tilde W+{\rm div}\, A+|A|^2$,
$$
\int_{{\Bbb R}^d}|A(x)|\,|x|^{1-d}dx<\infty, \quad \text{and}\quad
\int_{{\Bbb R}^d}|\tilde W|\,|x|^{1-d}dx<\infty
$$  due to \eqref{47} and \eqref{48}.
The proof is completed.

\section{Absolute continuity of the spectum for potentials of a special form}

According to Theorem~\ref{3.1} proved in the  preceding section,   Theorem~\ref{main}  follows  from  the  statement   formulated  below.
\begin{theorem}\label{main*}
Let  $V$ be a  real-valued  bounded  measurable  function on ${\Bbb R}^d$ representable in   the form
\begin{equation}\label{Vrep}
V(x)=\tilde W(x)+{\rm div}\, A(x)+ |A(x)|^2,
\end{equation}
where  the vector potential $A:{\Bbb R}^d\rightarrow{\Bbb R}^d$  and   the function $\tilde W:{\Bbb R}^d\rightarrow{\Bbb R}$  satisfy the conditions
\begin{equation}\label{AWcond}
\begin{split}
A\in L^\infty_{\rm loc}({\Bbb R}^d,{\Bbb R}^d)\cap  {\mathcal H}^1_{\rm loc}({\Bbb R}^d,{\Bbb R}^d),\quad \tilde W\in L^\infty_{\rm loc}({\Bbb R}^d),\\
\int_{{\Bbb R}^d}\frac{ (|\tilde W(x)|+|A(x)|^2)}{|x|^{d-1}}dx<\infty.\,\,\,\,\,\,\,\,\,\,\,\,\,\,\,\,\,\,\,\,
\end{split}
\end{equation}
Assume that the negative spectrum of the operator $H=-\Delta+V$ consists  of  eigenvalues $\{\lambda_j\}$  obeying the condition
$$
\sum_j\sqrt{|\lambda_j|}<\infty.
$$
Then the absolutely continuous spectrum of the  operator $H=-\Delta+V$  is essentially supported  on $[0,\infty)$.
\end{theorem}

Theorem~\ref{3.1} is a consequence of  a certain estimate of the entropy of the spectral measure  corresponding to an element $f\in L^2({\Bbb R}^d)$. 
 This measure is defined  as  a unique non-negative measure $\mu$ on ${\Bbb R}$  having the   property
$$
\bigl((H-z)^{-1}f,f\bigr)=\int_{{\Bbb R}}\frac{d\mu(t)}{(t-z)},\qquad \forall z\in {\Bbb C}\setminus {\Bbb R}.
$$

\begin{theorem}
Let the conditions of Theorem~\ref{3.1}  be  fulfilled. Then
there is a vector  $f\in L^2({\Bbb R}^d)$ such that, for any $0<a<b<\infty$,
\begin{equation}\label{thm5inequality}
\int_{a}^b\log \bigl(\mu'(\lambda)\bigr)\lambda^{-1/2}d\lambda\geq  -C_d\Bigl(\int_{{\Bbb R}^d}(\tilde W+|A|^2)|x|^{1-d}dx+\sum_j\sqrt{|\lambda_j|}\Bigr)-
\alpha_d(a,b;\|V\|_\infty),
\end{equation}
where the  constant $C_d>0$ depends  only on   the  dimension $d$,  while $\alpha_d(a,b;\|V\|_\infty)$ depends on $a$, $b$, the dimension $d$ and the norm $\|V\|_\infty$.
\end{theorem}

If the right hand side of \eqref{thm5inequality}   is finite, then
 $\mu'(\l)>0$ for  almost every $\l>0$. Therefore \eqref{thm5inequality} implies Theorem~\ref{main*}.

An important part of  the proof of this theorem is related to approximations of  the spectral measure  of the operator $-\Delta+W+V$
by spectral measures  of similar operators with compactly supported potentials.
We  have to consider  several cases, one of  which is the case  where the potential is unbounded.
The operator in this case can be defined  in the sense of  quadratic forms.

Let us  recall certain facts  of this theory. Let $a[u,v] $  be a  closed semi-bounded  sesquilinear form in a Hilbert space ${\frak H}$.
Semi-boundedness means that
$$
a[u,u]\geq -C \|u\|^2,\qquad   \forall u\in {\rm Dom}\,[a],
$$
with some  positive constant $C>0$. Closedness means that for any $\tau>C$, the domain $ {\rm Dom}\,[a]$ of the form
is a complete Hilbert space with  respect to the inner product
$$
a[u,v]+\tau(u,v).
$$
There is a  unique self-adjoint operator $A$ corresponding to the form $a$, such that $ {\rm Dom}\,A \subset {\rm Dom}\,[a]$
and
$$
(A u,v)=a[u,v]\qquad \forall u,v \in {\rm Dom}\,[a].
$$
A  vector $u\in {\frak H}$  belongs to $ {\rm Dom}\,A$  if and  only if  there  is a  vector $w\in {\frak H}$
such that
$$
 a[u,v]=(w,v)\qquad \forall v\in  {\rm Dom}\,[a].
$$
In this case,
$A u=w $.

First  consider a  Schrodinger operator $-\Delta+\tilde W_- +V$,
where $V$ and $\tilde W_- \geq0$  obey the conditions \begin{equation}\label{53} V\in L^\infty({\Bbb R}^d),\quad
 \tilde W_-\in L^\infty_{\rm loc}({\Bbb R}^d),
\end{equation}
\begin{equation}\label{53b}
\int_{{\Bbb R}^d}\frac{\tilde W_-}{|x|^{d-1}}dx<\infty.
\end{equation}
We define $-\Delta+\tilde W_- +V$ as  the operator corresponding to the quadratic  form
$$
\int_{{\Bbb R}^d}\Bigl(|\nabla u|^2+(\tilde W_- +V)|u|^2\Bigr)dx.
$$
The domain of this quadratic  form consists of  all ${\mathcal H}^1({\Bbb R}^d)$-functions that are square 
integrable with respect to  the measure $\tilde W_-dx$.

\begin{proposition}\label{4.3}
Let $f\in L^2({\Bbb R}^d)$ and let $V$  and $\tilde W_-\geq0$  satisfy \eqref{53}.
Assume that $u\in {\rm Dom}\, (-\Delta+\tilde W_- +V)$ is a solution of the equation
$$
-\Delta u+(\tilde W_- +V-z)u=f,\qquad {\rm Im\,}z\neq 0.
$$
Then
$$
\|u\|_{{\mathcal H}^1}\leq C \|f\|_{L^2}
$$
with
$$
C=\sqrt{\Bigl((3/2+|{\rm Re}\, z|+\|V\|_\infty)/|{\rm Im}\, z|^2+1/2\Bigr)}.
$$
\end{proposition}

{\it Proof.}  Since
$$
\int_{{\Bbb R}^d}|\nabla u|^2dx+\int_{{\Bbb R}^d}(\tilde W_- +V-z)|u|^2dx=\int_{{\Bbb R}^d}f\bar udx,
$$
we conclude that
$$
\int_{{\Bbb R}^d}|\nabla u|^2dx+\int_{{\Bbb R}^d}(\tilde W_- +V-{\rm Re}\, z)|u|^2dx= {\rm Re}\,\int_{{\Bbb R}^d}f\bar udx,
$$
Consequently,
$$
\int_{{\Bbb R}^d}|\nabla u|^2dx\leq (1/2+\|V\|_\infty+|{\rm Re}\, z|)\int_{{\Bbb R}^d}|u|^2dx+ \frac12\int_{{\Bbb R}^d}|f|^2 dx.
$$
It remains to note that $\|u\|_{L^2}\leq ({\rm Im}\,z)^{-1}\|f\|_{L^2}$. $\,\,\,\,\,\,\,\,\,\,\,\,\,\Box$

\bigskip

Let  $V$ be a  real-valued  bounded  measurable  function on ${\Bbb R}^d$ representable in   the form
\eqref{Vrep},
where  the vector potential $A:{\Bbb R}^d\rightarrow{\Bbb R}^d$  and   the function $\tilde W:{\Bbb R}^d\rightarrow{\Bbb R}$  satisfy the conditions
\eqref{AWcond}.
Let  $\theta$  be a  smooth real-valued  function on ${\Bbb R}$  having the property
\begin{equation}\label{theta}
\theta(t)=\begin{cases}
1\qquad \text{if}\quad t<0,\\
0\qquad\text{if}\quad  t>1.
\end{cases}
\end{equation}
 For a natural number $n$, we define $\theta_n$  by
\begin{equation}\label{thetan}
\theta_n(x)=\theta(|x|-n),\qquad x\in {\Bbb R}^d.
\end{equation}
After that,  we set 
\begin{equation}\label{Vn=}
V_{n}=\theta_n (\tilde W_-+ V)+|\nabla\theta_n \cdot A|+\nabla\theta_n \cdot A-\chi_R \tilde W_-,
\end{equation}
where $\tilde W_-=\frac12(|\tilde W|-\tilde W)$  is the negative  part of the function $\tilde W$  and $\chi_R$  is the characteristic  function of the ball $\{x\in {\Bbb R}^d:\,\, |x|<R\}$.

Now, for a  fixed   function  $f\in L^2({\Bbb R}^d)$,  define the non-negative measures $\mu_n$  and $\mu$ on ${\Bbb R}$   by
\begin{equation}\label{mun}
\Bigl((-\Delta+V_{n}-z)^{-1}f,f\Bigr)=\int_{\Bbb R}\frac{d\mu_n(t)}{t-z},\qquad  \forall z\in {\Bbb C}\setminus {\Bbb R},
\end{equation}
and
\begin{equation}\label{mu}
\Bigl((-\Delta+(1-\chi_R)W_-+V-z)^{-1}f,f\Bigr)=\int_{\Bbb R}\frac{d\mu(t)}{t-z},\qquad  \forall z\in {\Bbb C}\setminus {\Bbb R}.
\end{equation}

\begin{proposition}\label{mun3.2} Let $\mu_n$  and $\mu$  be  the measures  defined  by \eqref{mun} and \eqref{mu}.
Then the sequence $\mu_n$  converges to $\mu$ in the weak-$*$   topology, i.e.
for any   compactly  supported  continuous   function $\phi\in C({\Bbb R})$,
$$
\int_{\Bbb R}\phi(t) \,d\mu_n(t)\to  \int_{\Bbb R}\phi(t) \, d\mu(t),\qquad \text{as}\quad n\to\infty.
$$
\end{proposition}

{\it Proof.} Since any compactly  supported  function  $\phi\in C({\Bbb R})$ can be  approximated   by  finite linear combinations of   functions of the   form $\phi_z(t)={\rm Im}\,\bigl(1/(t-z)\bigr)$,
 it is  sufficient to  show that
$$
\int_{\Bbb R}\frac{d\mu_n(t)}{t-z}\to \int_{\Bbb R}\frac{d\mu(t)}{t-z}\qquad \text{as}\quad n\to\infty,  \quad \forall z\in {\Bbb C}\setminus {\Bbb R},
$$
uniformly on compact sets in ${\Bbb C}\setminus {\Bbb R}$,
which is the same as showing that 
$$
\Bigl((-\Delta+V_n-z)^{-1}f,f\Bigr)\to \Bigl((-\Delta+(1-\chi_R)\tilde W_-+V-z)^{-1}f,f\Bigr)\qquad \text{as}\quad n\to\infty,  \quad \forall z\in {\Bbb C}\setminus {\Bbb R}.
$$
 Using  the Heine-Borel lemma, one can reduce it to one point $ z\in {\Bbb C}\setminus {\Bbb R}.$
In order to establish the required convergence,  we  use   Hilbert's identity   saying that
\begin{equation}\notag\begin{split}
\Bigl((-\Delta+V_n-z)^{-1}f,f\Bigr)- \Bigl((-\Delta+(1-\chi_R)\tilde W_-+V-z)^{-1}f,f\Bigr)=\\
\Bigl(((1-\chi_R)\tilde W_-+V-V_n)(-\Delta+V_n-z)^{-1}f,(-\Delta+(1-\chi_R)\tilde W_-+V-\bar z)^{-1} f\Bigr).
\end{split}
\end{equation}
It becomes  clear that
to prove  the  proposition, one  needs  to show that
$$
\int_{{\Bbb R}^d}((1-\chi_R)\tilde W_-+V-V_n)u_n(x)\bar u(x)\,dx\to 0,\qquad \text{as}\quad n\to\infty,
$$
where
$$
u_n=(-\Delta+V_n-z)^{-1}f\qquad \text{and}\quad u=(-\Delta+(1-\chi_R)\tilde W_-+V-\bar z)^{-1} f.
$$
Let us first establish the relation
\begin{equation}\label{58-1}
\int_{{\Bbb R}^d}(1-\theta_n)(\tilde W_-+V)u_n(x)\bar u(x)\,dx\to 0,\qquad \text{as}\quad n\to\infty.
\end{equation}
According to Proposition~\ref{4.3},
\begin{equation}\label{58}
\sup_n\|u_n\|_{{\mathcal H}^1}<\infty, \qquad \text{and}\qquad \|u\|_{{\mathcal H}^1}<\infty.
\end{equation}
On the other  hand, for $n>R$,
\begin{equation}\label{58b}\begin{split}
\int_{{\Bbb R}^d} \Bigl(   (-\nabla  \theta_n ) u_n+(1- \theta_n )\nabla  u_n \Bigr) \nabla \bar u \, dx+\\+
\int_{{\Bbb R}^d} (1-\theta_n(x))(\tilde W_-+V-z)u_n(x)\bar u(x)\,dx=
\int_{{\Bbb R}^d} (1-\theta_n(x)) u_n(x)f(x)\,dx.
\end{split}
\end{equation}
Thus \eqref{58-1} follows  from \eqref{58b} by \eqref{58}.

Since $(1-\chi_R)\tilde W_-+V-V_n=(1-\theta_n)(\tilde W_-+V)-|\nabla\theta_n \cdot A|-\nabla\theta_n \cdot A$, it remains to show that
\begin{equation}\label{581}
\int_{{\Bbb R}^d}(|\nabla\theta_n \cdot A|+\nabla\theta_n \cdot A) u_n(x)\bar u(x)\,dx\to 0,\qquad \text{as}\quad n\to\infty.
\end{equation}

Replacing $1-\theta_n$  by $(1-\theta_{n-1})\theta_{n}$ and  $-\nabla \theta_n$  by $\nabla (1-\theta_{n-1})\theta_{n}$  in \eqref{58b},
one can easily  show that\begin{equation}\label{582}
\int_{{\Bbb R}^d}(1-\theta_{n-1})\theta_{n}(\tilde W_-+V) u_n(x)\bar u(x)\,dx\to 0,\qquad \text{as}\quad n\to\infty.
\end{equation}
Using the  equality
\begin{equation}\notag\begin{split}
\int_{{\Bbb R}^d}  \nabla  u_n \Bigl(   (-\nabla  \theta_{n-1} ) \bar u+(1- \theta_{n-1} )\nabla  \bar u \Bigr) \, dx+\\
+\int_{{\Bbb R}^d} (1-\theta_{n-1}(x))(V_n-z)u_n(x)\bar u(x)\,dx=\int_{{\Bbb R}^d} (1-\theta_{n-1}(x))f(x)\bar u(x)\,dx,
\end{split}
\end{equation} one also obtains
\begin{equation}\label{583}
\int_{{\Bbb R}^d}(1-\theta_{n-1})V_n(x) u_n(x)\bar u(x)\,dx\to 0,\qquad \text{as}\quad n\to\infty.
\end{equation}
Since $V_n=\theta_n(\tilde W_-+V)+|\nabla\theta_n \cdot A|+\nabla\theta_n \cdot A-\chi_R \tilde W_-$ and $\nabla\theta_n=(1-\theta_{n-1})\nabla\theta_n$,  the relation \eqref{581} follows  from
\eqref{582} and \eqref{583}.
$\,\,\,\,\,\,\,\,\,\Box$

\bigskip

Let  $V$ be representable in   the form
\eqref{Vrep},
where $A:{\Bbb R}^d\rightarrow{\Bbb R}^d$  and    $\tilde W:{\Bbb R}^d\rightarrow{\Bbb R}$  satisfy the conditions
\eqref{AWcond}.
Let    $\chi_n$  be  the characteristic  function of the ball $\{x\in {\Bbb R}^d:\,\,|x|<R\}$.
This time,  we set 
$$
V_n=(1-\chi_n )\tilde W_-+ V,
$$
where $\tilde W_-=\frac12(|\tilde W|-\tilde W)$  is the negative  part of the function $\tilde W$.
For a  fixed   function  $f\in L^2({\Bbb R}^d)$,  define the non-negative measures $\mu_n$  and $\mu$ on ${\Bbb R}$   by
\begin{equation}\label{munb}
\Bigl((-\Delta+V_n-z)^{-1}f,f\Bigr)=\int_{\Bbb R}\frac{d\mu_n(t)}{t-z},\qquad  \forall z\in {\Bbb C}\setminus {\Bbb R},
\end{equation}
and
\begin{equation}\label{mub}
\Bigl((-\Delta+V-z)^{-1}f,f\Bigr)=\int_{\Bbb R}\frac{d\mu(t)}{t-z},\qquad  \forall z\in {\Bbb C}\setminus {\Bbb R}.
\end{equation}

\begin{proposition} \label{mun3.3} Let $\mu_n$  and $\mu$  be  the measures  defined  by \eqref{munb} and \eqref{mub}.
Then the sequence $\mu_n$  converges to $\mu$ in the weak-$*$   topology, i.e.
for any   compactly  supported  continuous   function $\phi\in C({\Bbb R})$,
$$
\int_{\Bbb R}\phi(t) \,d\mu_n(t)\to  \int_{\Bbb R}\phi(t) \, d\mu(t),\qquad \text{as}\quad n\to\infty.
$$
\end{proposition}

{\it Proof.}  It suffices to show that 
$$
\Bigl((-\Delta+V_n-z)^{-1}f,f\Bigr)\to \Bigl((-\Delta+V-z)^{-1}f,f\Bigr)\qquad \text{as}\quad n\to\infty,  \quad \forall z\in {\Bbb C}\setminus {\Bbb R}.
$$
In order to establish the required convergence,  we  use   Hilbert's identity   saying that
\begin{equation}\notag\begin{split}
\Bigl((-\Delta+V_n-z)^{-1}f,f\Bigr)- \Bigl((-\Delta+V-z)^{-1}f,f\Bigr)=\\
\Bigl((V-V_n)(-\Delta+V_n-z)^{-1}f,(-\Delta+V-\bar z)^{-1} f\Bigr).
\end{split}
\end{equation}
It becomes  clear that
to prove  the  proposition, one  needs  to show that
$$
\int_{{\Bbb R}^d}(V-V_n)u_n(x)\bar u(x)\,dx\to 0,\qquad \text{as}\quad n\to\infty,
$$
where
$$
u_n=(-\Delta+V_n-z)^{-1}f\qquad \text{and}\quad u=(-\Delta+V-\bar z)^{-1} f.
$$
Put  differently, we  have to establish the relation
\begin{equation}\label{58-1q}
\int_{{\Bbb R}^d}(1-\chi_n)\tilde W_- u_n(x)\bar u(x)\,dx\to 0,\qquad \text{as}\quad n\to\infty.
\end{equation}
According to Proposition~\ref{4.3},
\begin{equation}\label{58q}
\sup_n\|u_n\|_{{\mathcal H}^1}<\infty, \qquad \text{and}\qquad \|u\|_{{\mathcal H}^1}<\infty.
\end{equation}
On the other  hand,
\begin{equation}\label{58bq}\begin{split}
\int_{{\Bbb R}^d}  \nabla  u_n \Bigl(   (-\nabla  \theta_{n-1} ) \bar u+(1- \theta_{n-1} )\nabla  \bar u \Bigr) \, dx+\\
+\int_{{\Bbb R}^d} (1-\theta_{n-1}(x))(V_n-z)u_n(x)\bar u(x)\,dx=\int_{{\Bbb R}^d} (1-\theta_{n-1}(x))f(x)\bar u(x)\,dx,
\end{split}
\end{equation}
where $\theta_n$ are defined  by \eqref{theta} and \eqref{thetan}.
Combining \eqref{58bq} and \eqref{58q}, we obtain  that
$$\int_{{\Bbb R}^d} (1-\theta_{n-1}(x))V_n u_n(x)\bar u(x)\,dx\to 0,\qquad \text{as}\quad n\to\infty.$$
 The latter relation implies \eqref{58-1q}, because $(1-\chi_n)\tilde W_-= (1-\theta_{n-1})(V_n-V)$.
$\,\,\,\,\,\,\,\,\,\Box$

\bigskip

\section{The entropy of a measure}

Let $\mu$  be an arbitrary non-negative   finite Borel measure  on the real line ${\Bbb R}$.
It can be decomposed  into the sum of three terms
$$
\mu=\mu_{ac}+ \mu_{pp}+\mu_{sc}
$$
where the first term is absolutely continuous, the second term is  pure point, and the last term is singular continuous with respect to the Lebesgue measure.
The limit
$$
\mu'(\lambda)=\lim_{\varepsilon\to0}\frac{\mu(\lambda-\varepsilon,\lambda+\varepsilon)}{2\varepsilon}
$$
exists  and coincides  with $\mu'_{ac}(\lambda)$  for  almost every  $\lambda\in {\Bbb R}$.
Therefore  the fact  that $\mu'>0$  almost everywhere on ${\Bbb R}_+=[0,\infty)$ implies that  the support
 of the absolutely continuous part of the measure contains ${\Bbb R}_+$. A useful tool that often allows 
to  understand the structure of the set 
$$
\{\lambda\in {\Bbb R}:\,\,\mu'(\lambda)>0 \}
$$
is the  entropy  of  one measure with respect  to the other.

\bigskip

{\bf Definition.}  Let $\rho$  and $\nu$  be  finite  Borel measures on a  compact Hausdorff  space $X$.
We define the entropy of the measure $\rho$   relative to $\nu$ by
$$
S(\rho|\nu)=\begin{cases}
-\infty,\qquad\qquad \text{if}\quad \rho  \text{  is not }\,\, \nu-\text{ac}\\
-\int_X \log(\frac{d\rho}{d\nu})d\rho, \qquad \text{if}\quad \rho  \text{  is  }\,\,\nu-\text{ac}.
\end{cases}
$$

\bigskip

The following  result  was proved in the remarkable paper \cite{KS} by Killip and Simon.

\begin{theorem}\label{4.1}
The entropy is  jointly upper semi-continuous in $\rho$  and $\nu$  with respect to  the weak-$*$ topology.
That is,
 if $\rho_n\to\rho$ and $\nu_n\to \nu$ as $n\to\infty$, then
$$
S(\rho|\nu)\geq \limsup_{n\to\infty} S(\rho_n|\nu_n).
$$
\end{theorem}

\bigskip
The weak-$*$ convergence in this theorem means convergence of  the sequence  of integrals of an arbitrary continuous function on $X$ 
with respect  to  the measures $\rho_n$ and $\nu_n$. The definition of the weak-$*$ convergence of  measures  on ${\Bbb R}$
involves integrals of  continuous  functions on ${\Bbb R}$ which  can not be  viewed as a compact space $X$.

\begin{corollary}\label{4.2}
Let $\mu_n$ be a  sequence  of finite Borel measures on the real line ${\Bbb R}$  converging to a  finite  Borel measure $\mu$ in the weak-$*$ sense. That is
$$
\int_{\Bbb R}\phi(\lambda)d\mu_n(\lambda)\to \int_{\Bbb R}\phi(\lambda)d\mu_n(\lambda),\qquad \text{as}\quad n\to\infty,
$$
for any compactly supported  continuous  function $\phi$ on ${\Bbb R}$. Then  for any $0<a<b<\infty$,
\begin{equation}\label{65}
\int_a^b \log(\mu'(\lambda))\l^{-1/2}d\l\geq  \limsup_{n\to\infty} \int_a^b \log(\mu'_n(\lambda))\l^{-1/2}d\l.
\end{equation}
\end{corollary}

{\it Proof.} Choose $\ve>0$ so that $a-\ve>0$. Set $X=[a-\ve,b+\ve], \, d\rho=\chi_{[a,b]}(\l)\l^{-1/2}d\l$  and $d\nu_n=\theta(\l) d\mu_n$,
where $\theta$ is a continuous  function on  ${\Bbb R}$ vanishing  outside of $X$  and equal to 1 on $[a,b]$. 
The notation $\chi_{[a,b]}$  is  used   for the characteristic  function of the interval $[a,b]$.
Consider   $\rho$  and $\nu_n$  as  measures  on $X$. According to Theorem~\ref{4.1},
\begin{equation}\notag
\int_a^b \log\Bigl(\mu'(\lambda)\l^{1/2}\Bigr)\l^{-1/2}d\l\geq  \limsup_{n\to\infty} \int_a^b \log\Bigl((\mu'_n(\lambda)\l^{1/2}\Bigr)\l^{-1/2}d\l.
\end{equation}
This  inequality is  equivalent to \eqref{65}. $\,\,\,\,\,\,\,\Box$
\bigskip

\section{A "trace-type"  estimate for the spectral measure}
Let 
$T$ be the operator defined by 
\begin{equation}\label{Tef}
\bigl[Tu\bigr](r)=-\frac{d^2u}{dr^2}(r)+Q(r)u(r), \qquad r>1,
\end{equation}
where  $Q(r)$ is a selfadjoint $n\times n$-matrix  for  each $r>1$. The domain of $T$ consists of all  ${\mathcal H}^2([1,\infty);{\Bbb C}^n)$-functions   vanishing at the point $r=1$.
 We will assume  that $Q$ is a  continuous  compactly supported  function.

Let $e_0\in {\Bbb C}^n$ be the  vector whose  first  component is 1 and all other components are equal to zero.
Set
$f(r)=\chi_{[1,2]}(r) e_0$, where $\chi_{[1,2]}$ is  the characteristic  function of the interval $[1,2]$. We define the measure $\mu$
as the unique non-negative  measure on ${\Bbb R}$  obeying
\begin{equation}\label{measureT}
\Bigl((T-z)^{-1}f,f\Bigr)=\int_{{\Bbb R}}\frac{d\mu(t)}{t-z}\qquad  \text{ for all} \quad z\in {\Bbb C}\setminus {\Bbb R}.
\end{equation}

\begin{theorem}\label{2.1} Let $\{\lambda_j\}$ be the  negative eigenvalues of the operator \eqref{Tef}. Let $\mu$ be  defined by \eqref{measureT}. 
Assume that $$Q(r)e_0=0\qquad \text{ for all } \qquad r\leq 2.$$
Then for any  $0<a<b<\infty$,
\begin{equation}\label{35}\begin{split}
\int_a^b\log\Bigl(   \mu'(\lambda)\Bigr)\lambda^{-1/2}d\lambda+\int_a^b\log\Bigl(  \frac{(1-\cos(\sqrt{\lambda}))^2}{4\pi \lambda^{3/2}} \Bigr)\lambda^{-1/2}d\lambda\\\geq -\frac{\pi}2\int_2^\infty \Bigl(Q(r)e_0,e_0\Bigr)  dr-2\pi\sum_j\sqrt{|\lambda_j|}-2\pi \|Q_-\|_\infty^{1/2},
\end{split}
\end{equation}
where $Q_-(r)=\frac12(|Q(r)|-Q(r))$.
\end{theorem}
Except  for the replacement of $ \|Q\|_\infty$  by $ \|Q_-\|_\infty$,
the proof of \eqref{35} repeats word  by word the proof of Theorem 2.1  from my paper \cite{AHP}.
It was  overlooked in \cite{AHP}  that the bottom of the spectrum of a Schr\"odinger operator  with the potential $Q$ can be estimated  by $ \|Q_-\|_\infty$ instead of $ \|Q\|_\infty$.    

Let \begin{equation}\label{TH}\tilde H=-\Delta+V\end{equation}
  be the operator on $L^2({\Bbb R}^d\setminus B_1)$   with the Dirichlet  condition on the boundary of
the unit ball $B_1=\{x\in{\Bbb R}^d:\,\,\,\,|x|\leq 1\}$. Let $f(x)=|{\Bbb S}|^{-1/2}\chi_{[1,2]}(|x|)|x|^{-(d-1)/2}$  for all $x\in {\Bbb R}^d\setminus B_1$,
where $\chi_{[1,2]}$  is the characteristic function of the interval $[1,2]$ and $|{\Bbb S}|$ is the  area  of the unit sphere in ${\Bbb R}^d$.
We define the measure $\tilde \mu$
as the unique non-negative  measure on ${\Bbb R}$  obeying 
\begin{equation}\label{measureTH}
\Bigl((\tilde H-z)^{-1}f,f\Bigr)=\int_{{\Bbb R}}\frac{d\tilde \mu(t)}{t-z}\qquad  \text{ for all} \quad z\in {\Bbb C}\setminus {\Bbb R}.
\end{equation}

\begin{corollary}\label{c5.1} Let $V$ be a    continuous  real-valued    function on $\{x\in{\Bbb R}^d:\,\,\,\,|x|\geq 1\}$
having the property
$$
V(x)=\frac{-(d-1)(d-3)}{4|x|^2}\qquad  \text{for}\quad 1\leq |x|\leq 2.
$$
Let $\tilde \mu$  be defined  by \eqref{measureTH}  where $\tilde H$  is  the operator  defined  by \eqref{TH}. Assume that $V$ is compactly supported.
Then  for any  $0<a<b<\infty$,
\begin{equation}\label{35b}\begin{split}
\int_a^b\log\Bigl(   \tilde \mu'(\lambda)\Bigr)\lambda^{-1/2}d\lambda+\int_a^b\log\Bigl(  \frac{(1-\cos(\sqrt{\lambda}))^2}{4\pi \lambda^{3/2}} \Bigr)\lambda^{-1/2}d\lambda\\\geq -\frac{\pi}{2|{\Bbb S}|}\int_{|x|>2} \frac{V(x)}{|x|^{d-1}} dx-2\pi\sum_j\sqrt{|\lambda_j|}-2\pi \Bigl(\|V_-\|_\infty+\frac14 \Bigr)^{1/2}-\frac{(d-1)(d-3)}8,
\end{split}
\end{equation}
where $V_-(x)=\frac12(|V(x)|-V(x))$  and $|{\Bbb S}|$ is the  area  of the unit sphere in ${\Bbb R}^d$.
\end{corollary}

{\it Proof.} Let $r$  and $\theta$  be the polar coordinates in ${\Bbb R}^d$. For each natural number $n$, we define $P_n$ to  be the  orthogonal projection in $L^2({\Bbb R}^d\setminus B_1)$
onto  the space of  functions  of the  form $v(r)Y_n(\theta)$, where $Y_n(\theta)$  is the  $n$-th eigenfunction of the Laplace-Beltrami operator $-\Delta_\theta$ on the  unit  sphere.
Define also $\tilde P_n$  by
$$
\tilde P_n=\sum_{j=1}^n P_j.
$$
Then $\tilde P_n\to I$  strongly as $n\to\infty$. Using this property, one can easily  show that
$$
\tilde P_nV\tilde P_n u\to V u\qquad \text{ as }\quad n\to\infty\qquad \text{  for each} \quad u\in L^2({\Bbb R}^d\setminus B_1).
$$
Consequently,
$$
\Bigl( (-\Delta  + \tilde P_nV\tilde P_n-z)^{-1}f,f\Bigr)\to  \Bigl( (-\Delta +V-z)^{-1}f,f\Bigr),\qquad \text{as}\quad n\to\infty,
$$
for each $z\in {\Bbb C}\setminus {\Bbb R}$ and $f\in L^2({\Bbb R}^d\setminus B_1)$. This  relation implies the  weak-$*$  convergence of  the  corresponding spectral measures:
\begin{equation}\label{mutotilde}
\tilde \mu_n\to \tilde \mu,\qquad \text{as}\quad n\to \infty,
\end{equation}
where $\tilde \mu_n$  is defined by
$$
\Bigl((-\Delta  + \tilde P_nV\tilde P_n-z)^{-1}f,f\Bigr)=\int_{{\Bbb R}}\frac{d\tilde \mu_n(t)}{t-z}\qquad  \text{ for all} \quad z\in {\Bbb C}\setminus {\Bbb R}.
$$
On the other hand, the measure $\tilde \mu_n$ coincides  with the spectral measure \eqref{measureT} of the operator \eqref{Tef} with the potential  $Q$
$$
Q= \tilde P_n V\tilde P_n +\frac{(d-1)(d-3)}{4r^2}\tilde P_n-\frac{1}{r^2}\Delta_\theta \tilde P_n.
$$
This matrix-valued potential $Q$ can be also  approximated  by compactly supported  matrix-valued potentials  $$Q_l=\chi_{[1,l]}(r)Q,$$ where $\chi_{[1,l]}$ is the characteristic  function of the interval $[1,l]$.
Since 
$$
\|Q-Q_l\|_\infty\to0,\qquad \text{as}\quad l\to\infty.
$$
we obtain  that
$$
\Bigl( (-d^2/dr^2  +Q_l-z)^{-1}f,f\Bigr)\to  \Bigl( (-d^2/dr^2 +Q-z)^{-1}f,f\Bigr),\qquad \text{as}\quad l\to\infty,
$$
for each $z\in {\Bbb C}\setminus {\Bbb R}$. Therefore  the sequence of  the corresponding measures  $\nu_l$  constructed  for the operators $-d^2/dr^2  +Q_l$
converges to $\tilde \mu_n$  in the weak-$*$ topology  as $l\to\infty$.
 Theorem~\ref{2.1}  tells us that
\begin{equation}\notag \begin{split}
\int_a^b\log\Bigl(   \tilde \nu'_l(\lambda)\Bigr)\lambda^{-1/2}d\lambda+\int_a^b\log\Bigl(  \frac{(1-\cos(\sqrt{\lambda}))^2}{4\pi \lambda^{3/2}} \Bigr)\lambda^{-1/2}d\lambda\\\geq -\frac{\pi}{2|{\Bbb S}|}\int_{|x|>2} \frac{V(x)}{|x|^{d-1}} dx-2\pi\sum_j\sqrt{|\tilde \lambda_j|}-2\pi \Bigl(\|V_-\|_\infty+\frac14\Bigr)^{1/2}-\frac{(d-1)(d-3)}8,
\end{split}
\end{equation}
where $\tilde \lambda_j$ are  the negative eigenvalues  of the operator
$
-d^2/dr^2  +Q_l.
$
Hence,  by Corollary~\ref{4.2}, the  following  inequality  holds  for  the measure $\tilde \mu_n$  and  the negative eigenvalues
$\{ \Lambda_j\}$
 of   the operator
$-\Delta  + \tilde P_nV\tilde P_n$:
\begin{equation}\label{PnVPnmu} \begin{split}
\int_a^b\log\Bigl(   \tilde \mu'_n(\lambda)\Bigr)\lambda^{-1/2}d\lambda+\int_a^b\log\Bigl(  \frac{(1-\cos(\sqrt{\lambda}))^2}{4\pi \lambda^{3/2}} \Bigr)\lambda^{-1/2}d\lambda\\\geq -\frac{\pi}{2|{\Bbb S}|}\int_{|x|>2}\frac{V(x)}{|x|^{d-1}} dx-2\pi\sum_j\sqrt{| \Lambda_j|}-2\pi \Bigl(\|V_-\|_\infty+\frac14\Bigr)^{1/2}-\frac{(d-1)(d-3)}8,
\end{split}
\end{equation}
  Using Corollary~\ref{4.2}    one more time,
we infer \eqref{35b}  from \eqref{mutotilde} and \eqref{PnVPnmu}. $\,\,\,\,\,\,\,\,\,\,\Box$

\section{Eigenvalue sums stay bounded}

Let $V_n$  be the sequence of potentials  defined by \eqref{Vn=}. Assume that $A(x)=0$   for $|x|<2$.  Then
$$
\int_{|x|>2}|x|^{1-d}V_ndx=\int_{|x|>2}|x|^{1-d}\Bigl(\theta_n(\tilde W_++|A|^2)+|\nabla\theta_n \cdot A|-\chi_R \tilde W_-\Bigr)dx,
$$
where $\tilde W_+=\frac12(|\tilde W|+\tilde W)$  is the positive ppart of $\tilde W$. Consequently,  
$$
\int_{|x|>2}|x|^{1-d}V_ndx\leq \int_{{\Bbb R}^d}|x|^{1-d}\Bigl(|\tilde W|+|A|^2\Bigr)dx+c\bigl(\int_{|x|>n}|x|^{1-d}|A|^2dx\bigr)^{1/2}
$$
with some universal constant $c>0$. It is also easy to see that $\|(V_n)_-\|_\infty\leq \|V_-\|_\infty$  for $n>R$.
It is  more difficult to  prove that the eigenvalue sums $\sum_j|\l_j(V_n)|^{1/2}$ for  the operators $-\Delta +V_n$ have an upper  bound  independent  of  $n$.
This fact follows  from the  proposition stated below.

\begin{proposition}\label{6.1}
There are numbers $N\in {\Bbb N}$ and $C>0$  such that  each
 operator $-\Delta+V_n$  has at most $N$   negative   eigenvalues  $\{\l_j(V_n)\}$  and  all of them obey the condition
$
|\l_j(V_n)|\leq C.
$
\end{proposition}

{\it Proof.} The quadratic  form of the operator $-\Delta+V_n$ can be estimated  from below  by
the  functional
$$
\int_{{\Bbb R}^d}|\nabla u-\theta_n A u|^2dx-\int_{{\Bbb R}^d}\chi_R \tilde W_-(x)|u|^2dx,\qquad u\in {\mathcal H}^1({\Bbb R}^d).
$$
For $n>R$, the value of this funcional at  $u\in {\mathcal H}^1({\Bbb R}^d)$ does not exceed
$$
\int_{B_R}|\nabla u- A u|^2dx-\int_{B_R}\tilde W_-(x)|u|^2dx,
$$
where $B_R=\{x\in {\Bbb R}^d:\,\,|x|<R\}$ is  the ball of radius $R>0$ centered at the origin. 

Since
$$
\int_{B_R}|\nabla u- A u|^2dx\geq \int_{B_R}\Bigl(\frac12|\nabla u|^2- |A u|^2\Bigr)dx, 
$$
we  conclude that the iegenvalues of $-\Delta+V_n$  can be  estimated   from below by eigenvalues of the operator  $-\Delta/2-|A|^2-\tilde W_-$  on the ball $B_R$.
It remains to note that the spectrum of  the latter operator   is  discrete and semi-bounded. $\,\,\,\,\,\,\Box$

\begin{proposition}\label{6.2}
Both Propositions ~\ref{mun3.2}  and \ref{mun3.3}  hold in the case where the operator $-\Delta$  on ${\Bbb R}^d$
is replaced  by  the operator $-\Delta$   on  the domain  ${\Bbb R}^d\setminus B_1$  with the Dirichlet  boundary conditions on the unit sphere.
\end{proposition}

{\it Proof.} The arguments  used  in the proofs of  Propositions ~\ref{mun3.2}  and \ref{mun3.3}  are  suitable  for the operators on ${\Bbb R}^d\setminus B_1$. $\,\,\,\,\,\,\Box$

\bigskip

\begin{corollary} \label{6.3}
Let  $V$ be a  real-valued  measurable  function on ${\Bbb R}^d$ representable in   the form
\begin{equation}\label{Vrep*}
V(x)=(1-\chi_R)\tilde W_-(x)+\tilde W(x)+{\rm div}\, A(x)+ |A(x)|^2,
\end{equation}
where  the vector potential $A:{\Bbb R}^d\rightarrow{\Bbb R}^d$  and   the function $\tilde W:{\Bbb R}^d\rightarrow{\Bbb R}$  satisfy the conditions
\begin{equation}\label{AWcond*}
\begin{split}
A\in L^\infty_{\rm loc}({\Bbb R}^d,{\Bbb R}^d)\cap  {\mathcal H}^1_{\rm loc}({\Bbb R}^d,{\Bbb R}^d),\quad \tilde W\in L^\infty_{\rm loc}({\Bbb R}^d),\\
\int_{{\Bbb R}^d}\frac{ (|\tilde W(x)|+|A(x)|^2)}{|x|^{d-1}}dx<\infty.\,\,\,\,\,\,\,\,\,\,\,\,\,\,\,\,\,\,\,\,
\end{split}
\end{equation}
Assume that $A(x)=0$  for $|x|<2$  and  that
$$
\tilde W(x)=\frac{-(d-1)(d-3)}{4|x|^2}\qquad  \text{for}\quad 1\leq |x|\leq 2.
$$
Let $\tilde \mu$  be defined  by \eqref{measureTH}  where $\tilde H$  is  the operator  defined  by \eqref{TH}.   Finally, let $\{\l_j\}$  be the negative eigenvalues of $\tilde H$.
Then  for any  $0<a<b<\infty$,
\begin{equation}\label{35b*}\begin{split}
\int_a^b\log\Bigl(   \tilde \mu'(\lambda)\Bigr)\lambda^{-1/2}d\lambda+\int_a^b\log\Bigl(  \frac{(1-\cos(\sqrt{\lambda}))^2}{4\pi \lambda^{3/2}} \Bigr)\lambda^{-1/2}d\lambda\\\geq -\frac{\pi}{2|{\Bbb S}|}\int_{|x|>2} \frac{|\tilde W(x)|+|A(x)|^2}{|x|^{d-1}} dx-2\pi\sum_j\sqrt{|\lambda_j|}-2\pi \Bigl(\|V_-\|_\infty+\frac14 \Bigr)^{1/2}-\frac{(d-1)(d-3)}8,
\end{split}
\end{equation}
where $V_-(x)=\frac12(|V(x)|-V(x))$  and $|{\Bbb S}|$ is the  area  of the unit sphere in ${\Bbb R}^d$.
\end{corollary}
{\it Proof.} This  statement is a consequence of Corollaries ~\ref{4.2}, ~\ref{c5.1}  and  Propositions~\ref{6.1}, \ref{6.2}. $\,\Box$

\bigskip

\begin{theorem} \label{6.4}
Let  $V$ be a  real-valued  measurable  function on ${\Bbb R}^d$ representable in   the form
\begin{equation}\label{Vrep**}
V(x)=\tilde W(x)+{\rm div}\, A(x)+ |A(x)|^2,
\end{equation}
where  the vector potential $A:{\Bbb R}^d\rightarrow{\Bbb R}^d$  and   the function $\tilde W:{\Bbb R}^d\rightarrow{\Bbb R}$  satisfy the conditions
\eqref{AWcond*}.
Assume that \begin{equation}\label{AtildeW}
A(x)=0\qquad \text{  for} \,\,|x|<2, \quad\text{and  that}
\qquad
\tilde W(x)=\frac{-(d-1)(d-3)}{4|x|^2}\qquad  \text{for}\quad 1\leq |x|\leq 2.
\end{equation}
Let $\tilde \mu$  be defined  by \eqref{measureTH}  where $\tilde H$  is  the operator  defined  by \eqref{TH}  with $V$  representable in the form \eqref{Vrep**}.   Finally, let $\{\l_j\}$  be the negative eigenvalues of $\tilde H$.
Then  for any  $0<a<b<\infty$,
\begin{equation}\label{35b**}\begin{split}
\int_a^b\log\Bigl(   \tilde \mu'(\lambda)\Bigr)\lambda^{-1/2}d\lambda+\int_a^b\log\Bigl(  \frac{(1-\cos(\sqrt{\lambda}))^2}{4\pi \lambda^{3/2}} \Bigr)\lambda^{-1/2}d\lambda\\\geq -\frac{\pi}{2|{\Bbb S}|}\int_{|x|>2} \frac{|\tilde W(x)|+|A(x)|^2}{|x|^{d-1}} dx-2\pi\sum_j\sqrt{|\lambda_j|}-2\pi \Bigl(\|V_-\|_\infty+\frac14 \Bigr)^{1/2}-\frac{(d-1)(d-3)}8,
\end{split}
\end{equation}
where $V_-(x)=\frac12(|V(x)|-V(x))$  and $|{\Bbb S}|$ is the  area  of the unit sphere in ${\Bbb R}^d$.
\end{theorem}

{\it Proof.}  This theorem follows  from Corollary ~\ref{4.2},  Proposition~\ref{6.2} and Corollary ~\ref{6.3}. The inequality \eqref{35b**} is obtained by passing to the upper limit as $R\to\infty$ on both sides of \eqref{35b*}.  One only needs to   observe that  negative egenvalues  of  the Schr\"odinger  operator 
with the potential $(1-\chi_R)\tilde W_-+V$  are monotone functions of $R$. Therefore  they lie higher  than  negative  eigenvalues  of   the operator with the potential $V$.
$\,\,\,\,\,\Box$

\bigskip

Let $H=-\Delta+V$ be the Schr\"odinger operator  on the whole space ${\Bbb R}^d$ with an arbitrary  bounded  potential of the  form \eqref{Vrep}.
Assume that $A$ and $\tilde W$  obey \eqref{AWcond*}.
Define the function $\tilde V$ by
$$
\tilde V(x)=\frac{-(d-1)(d-3)\theta_2(x)}{4|x|^2}+(1-\theta_2(x)\tilde W(x)+{\rm div}\,\bigl( \theta_2(x)A(x)\bigr)+ |\theta_2(x)A(x)|^2,
$$
where $\theta_2$ is defined by  \eqref{theta} and  \eqref{thetan}  with $n=2$.
After  that, consider the operator $H_1=-\Delta+\tilde V$  on ${\Bbb R}^d\setminus B_1$  with the Dirichlet  boundary conditions on the unit sphere.
Since $\tilde V$ satisfies the conditions of Theorem~\ref{6.4}  imposed on $V$, an inequaity of the form \eqref{35b**}  holds  for  the spectral measure
of the operator $H_1$  corresponding to  some $f\in L^2({\Bbb R}^d\setminus B_1)$  that belongs to the absolutely  continuous subpace for the operator $H_1$.
By rather standard arguments of scattering theory, the absolutely continuous parts of  operators $H$ and $H_1$  are unitary equaivalent.
Therefore \eqref{thm5inequality}  also holds  with $C_d=\pi/(2|{\Bbb S}|)+2\pi$  and 
$$\alpha_d(a,b,\|V_-\|_\infty)=2\pi \Bigl(\|V_-\|_\infty+\frac14 \Bigr)^{1/2}+\frac{(d-1)(d-3)}8+\int_a^b\log\Bigl(  \frac{(1-\cos(\sqrt{\lambda}))^2}{4\pi \lambda^{3/2}} \Bigr)\lambda^{-1/2}d\lambda$$  for some $f\in L^2({\Bbb R}^d)$.


\begin{thebibliography}{99}


\bibitem{DR} Damanik, D., Remling, C.: {\it Schrödinger operators with many bound states}. Duke Math. J. {\bf 136}, 51--80 (2007)

\bibitem{KMS} Killip, R., Molchanov, S., Safronov, O.: {\it A relation between the positive and negative spectra of elliptic operators}. Lett. Math. Phys. {\bf  107},  (10), 1799--1807 (2017)

\bibitem{KS} Killip, R., Simon, B.:  {\it  Sum rules for Jacobi matrices and their applications to spectral theory}. Annals Math. {\bf 158} (1), 253 (2003)

\bibitem{JFA} Safronov, O.: {\it Multi-dimensional Schrödinger operators with some negative spectrum}. J. Funct. Anal. {\bf 238} (1), 327--339 (2006)

\bibitem{AHP}  Safronov, O.: {\it Absolutely Continuous Spectrum of a Dirac Operator in the Case of a Positive Mass}. Ann. Henri Poincare  {\bf 18}, 1385--1434 (2017)
\end{thebibliography}
\end{document}